\documentclass[aps,twocolumn,showpacs,nofootinbib,groupedaddress]
{revtex4}
\usepackage{epsf,amsmath,amssymb}
\usepackage{graphicx}% Include figure files
\newcommand{\Rsub}{\rm\scriptscriptstyle}
\begin{document}
%\thispagestyle{empty}
%\begin{center}
\title{Mass spectra of doubly heavy $\Omega_{QQ'}$ baryons}
%\vspace*{5mm}
\author{V.V.Kiselev}
\email{kiselev@th1.ihep.su}
\affiliation{State Research Center "Institute for High Energy Physics" \\
{Protvino, Moscow region, 142280 Russia}\\
Fax: +7-095-2302337}
\author{A.K.Likhoded,}
\affiliation{State Research Center "Institute for High Energy Physics" \\
{Protvino, Moscow region, 142280 Russia}\\
Fax: +7-095-2302337}
\author{O.N.Pakhomova}
\affiliation{Samara State University, Ac.Pavlov 1, Samara, 443011 Russia}
\author{V.A.Saleev}
\affiliation{Samara State University, Ac.Pavlov 1, Samara, 443011 Russia}
\begin{abstract}
{We evaluate the masses of baryons composed of two heavy quarks and a strange
quark with account for spin-dependent splittings in the framework of potential
model with the K$^2$O potential \cite{KKO} motivated by QCD with a three-loop
$\beta$ function for the effective charge consistent with both the perturbative
limit at short distances and linear confinement term at long distances between
the quarks. The factorization of dynamics is supposed and explored in the
nonrelativistic Schr\"odinger equation for the motion in the system of two
heavy quarks constituting the doubly heavy diquark and the strange quark
interaction with the diquark. The limits of approach, its justification and
uncertainties are discussed. Excited quasistable states are classified by the
quantum numbers of heavy diquark composed by the heavy quarks of the same
flavor.}
\end{abstract}

%\vspace*{1cm}
\pacs{14.20.Lq, 14.20.Mr, 12.39.Jh}

\maketitle
%\newpage
%\setcounter{page}{1}
\section{Introduction}

The long-lived doubly heavy baryons open a new battlefield in the study of
heavy quark dynamics in various aspects such as the interplay of strong and
weak interactions in decays as well as the confinement (see recent review on
this subject in \cite{revQQq}). Indeed, there are some features of these
baryons in contrast to the mixed flavor meson $B_c$ composed by two heavy
quarks, for which one can explore the technique of nonrelativistic QCD (NRQCD)
\cite{NRQCD} in the form of potential NRQCD (pNRQCD) \cite{pNRQCD} or
velocity-counting NRQCD (vNRQCD) \cite{LMR,MS,HMS}, while the description of
doubly heavy baryons involves the interaction with the light or strange quarks
constituting the baryon together with the doubly heavy diquark. Therefore, one
should combine the above NRQCD approaches with the heavy quark effective theory
(HQET) \cite{HQET} constructed in order to take into account the interactions
of local heavy spinor field with the soft light quarks and gluons. Thus, the
doubly heavy baryons give the possibility to test these methods in different
conditions specific for the baryons $QQq$ or $QQs$. 

An experimental opportunity for such the studies could be quite real at RunII
of Tevatron. Indeed, the $B_c$ meson was observed by the CDF Collaboration at
FNAL \cite{cdf} under a lower statistics of RunI. Its spectroscopic
characteristics and decays were theoretically described and predicted in the
framework of QCD sum rules \cite{QCDSRBc}, potential models \cite{PMBc,PRDBc}
and operator product expansion \cite{OPEBc}. Those predictions served to
isolate some reasonable constraints preferable for the detection of $B_c$
signal such as the intervals of mass, production rate, lifetime and branching
ratios of various decay modes with quite a high efficiency. The strategy on the
study of doubly heavy baryons follows the same way as for $B_c$ \cite{BTeV}.
First, one estimated the spectroscopic characteristics in the framework of
potential models \cite{PMQQq,PM1} and sum rules of QCD or NRQCD
\cite{QCDSRQQq}, that resulted in definite predictions of mass spectra and wave
functions, which determine the normalization of cross-sections in terms of soft
binding factor \cite{prodQQq}. Second, corresponding yields in hadronic
collisions and $e^+e^-$ annihilation can be obtained in the perturbative theory
(fourth order in $\alpha_s$ for the production of two pairs of heavy quarks in
QCD), so that the rates are large enough in order to expect a positive result
of experimental search for the doubly heavy baryons at RunII  \cite{prodQQq}.
Third, the potential models and operator product expansion were used to
calculate both exclusive and inclusive branching ratios \cite{PMdecay,OPEQQq},
respectively, as well as the lifetimes \cite{OPEQQq}. A new field of interest
for the investigations is radiative, electromagnetic or hadronic, transitions
between the quasistable states in the families of baryons with two heavy
quarks. A first step in the study of this problem was recently done in
\cite{guo2}, wherein some preliminary results were obtained on the
electromagnetic transitions between the levels of $\Xi_{bc}$.

In the present paper we analyze the basic spectroscopic characteristics for the
families of doubly heavy strange baryons $\Omega_{QQ'}=(QQ's)$ in the framework
of potential model and compare the results with the calculations in NRQCD sum
rules and lattice simulations for the ground states.

A general approach of potential models to calculate the masses of baryons
containing two heavy quarks was in detail discussed in \cite{PMQQq,PM1}. There
are two clear physical arguments in this problem, which differ by kinds of
interquark forces in the baryon. The first motivation on the form of
description is the pair interactions, while the second is the string-like
presentation based on the Wilson loop for three static sources, where one
source interacts with other two sources through a string connected to the
diquark. For the doubly heavy baryons, these two approaches result in quite
certain predictions, which can be clearly distinguished in the splitting
between the excitations as described by Gershtein et al. in \cite{PM1}. The
peculiarity of doubly heavy diquark is its small size, that can be used in
order to construct an effective approximation in the form of combined NRQCD and
HQET by implementing the hierarchy of following scales: the size of
$QQ'$-diquark antitriplet color subsystem, $r_{QQ'}$, which is about inverse
both the characteristic momentum transfer between the heavy quarks
$|{\boldsymbol k}|$ and the relative momentum of heavy quark motion
$|{\boldsymbol p}|\sim |{\boldsymbol k}| \sim m_{Q,Q'}\cdot v$ with a small
nonrelativistic velocity $v\ll 1$, the confinement scale $\Lambda_{QCD}$ for
the dynamics of light or strange quark and the heavy quark masses $m_{Q,Q'}$,
so that
$$
r_{QQ'}\cdot \Lambda_{QCD} \ll 1, \;\;\;\; \Lambda_{QCD}\ll m_Q.
$$
Under such conditions, we can factorize the doubly heavy diquark as a local
source of colored field in the interactions with the light quark, while the
dynamics in the diquark can be reliably described in terms of nonrelativistic
spinors, i.e. NRQCD, pNRQCD or vNRQCD. In the present paper we explore the
quark-diquark factorization of baryon wave-functions in the framework of
nonrelativistic potential model and use the QCD-motivated potential combining
the known perturbative calculations for the static potential at short distances
\cite{Peter,Schroed} and the linear confining term at large distances in the
way similar to the Buchm\"uler--Tye method \cite{BT} extended to the three-loop
$\beta$ function for the effective charge in the static potential \cite{KKO}.
Such the calculations differ from the approaches of pNRQCD and vNRQCD, since
the pNRQCD deals with the static potential of heavy sources fixed at a distance
$r$ not including the static energy of ultrasoft fields such as the
contributions from the quark-gluon string or sea, while the vNRQCD describes
the nonrelativistic coulomb system in the limit of negligible contributions by
ultrasoft fields valid at $m_{Q,Q'} \cdot v^2 \gg \Lambda_{QCD}$, that is
broken for the heavy-heavy systems composed of bottom and charmed quarks under
study. The static potential K$^2$O \cite{KKO} is consistent with the
high-virtuality normalization of coupling constant in QCD as well as with the
slope of Regge trajectories, that define two scale parameters of model. The
heavy quarkonia spectra \cite{KKO} and leptonic constants \cite{KLPS} are
described with quite a good accuracy in this approach, that results in the
fixed values of heavy quark masses implemented in the potential model with the
nonrelativistic quarks. The connection of these quantities with the pole and
current masses of heavy quarks was discussed in \cite{KKO}.

We adjust the nonrelativistic Schr\"odinger equation for the description of
light and strange quark dynamics. So, we determine the constituent mass of
light and strange quarks in the way of minimal binding energy of quarks in the
static field of colored source in the framework of presentation with the
constituent mass formed as a piece of ultrasoft energy in the linear confining
term of the potential or the string (see Section II.B). Then we test the
estimated mass values of light and strange quarks in the potential model by
calculating the masses of bound states with a single heavy quark, i.e. the
charmed and beauty mesons. We discuss uncertainties and apply the procedure for
the doubly heavy baryons.

If the heavy quarks composing the antitriplet-color diquark, have the same
flavor, $Q=Q'$, we have to take into account the Pauli principle for the
identical fermions. Then we find that for the symmetric, spatial parity P-even
wave functions of diquark, $\Psi_d({\bf r})$ (the orbital momentum equals $L_d=
2n$, where $n=0,1,2\ldots$) the spin wave-function of diquark should be
symmetric too, and the summed spin of quarks equals $S=1$, while for the
antisymmetric, -odd functions $\Psi_d({\bf r})$ (i.e. $L_d= 2n+1$), we have
$S=0$. 

Under factorization of diquark and strange quark dynamics, we accept the
following notations for the classification of levels in the system of $QQs$
baryons: the summed spin of two heavy quarks, their orbital momentum and
principal quantum number are denoted by the capital letters $S$, $L$ and $n_d$,
respectively, so that the ground state of diquark, for instance, is marked as
$n_d^{2S+1}L=1^3S$, while the motion of strange quark is labelled by a pair of 
lowercase letters $n_s\, l$. Such the notations are based on the presentation
that these quantum numbers are approximately conserved if we neglect the
multipole interactions in QCD \cite{12} with the emission of soft gluons, which
can be absorbed by the associated strange quark in the baryon or involved into
the scattering with the emission of kaon, for example, $g+s\to q+K$ with
$q=u,\, d$. In this expansion there are the following suppression factors
mentioned above: the diquark size with respect to the scale of confinement, the
relative momentum of heavy quarks with respect to the heavy quark mass, that
leads to suppression of both the chromoelectric and chromomagnetic dipole
transitions, respectively. Moreover, in the doubly heavy baryons with identical
flavors of two heavy quarks some transitions between the levels or mixing
operators have a double suppression, since for the corresponding operators one
could need the properties providing the change of summed heavy quark spin
$\Delta S =1 $ together with the change of their orbital momentum $\Delta L = 2
m+1$, $m\in {\mathbb Z}$. Say, we expect that the excited $2^1P$ level of
diquarks $bb$ and $cc$ would be quasistable under the transition to the ground
level $1^3S$. This picture can be not valid for the $bc$ diquark, since both
values of summed heavy quark spin are admissible, and the spin changing
operators are not removed and can result in a significant mixing of diquark
states labelled by the summed spin. So, we restrict our consideration of
$\Omega_{bc}$ mass spectrum by the spin-dependent splitting of ground state,
while for the doubly heavy baryons $\Omega_{cc}$ and $\Omega_{bb}$ we present
complete spectra of families.

Next point is the following evident condition for the applicability of diquark
dynamics factorization. So, we calculate the size of diquark subsystem in order
to demonstrate the reliability of calculations. We will see that the low-lying
levels have quite small sizes, especially in the case of $bb$ diquark, while
the higher excitations are large with respect to the distance to the strange
quark, so that the approximation of local colored source for such diquark
levels results in less reliable estimates of baryon masses.

The paper is organized as follows. In Section II we describe the static
potential explored in the paper and calculate the relevant constituent masses
of light and strange quarks. We test the accepted approximations in estimates
of masses for mesons with a single heavy quark. Then we introduce the
spin-dependent forces. In Section III we present numerical results. Our
conclusions are given in Section IV. 

\section{Nonrelativistic potential model}

Under the factorization of dynamics inside the doubly heavy diquark and in the
system of strange quark and diquark, we use the nonrelativistic Schr\"odinger
equation in order to solve the corresponding two-body problems, that yields
spin-independent wave-functions and levels. We use the three-loop improved
static potential in the Schr\"odinger equations with the heavy quark masses
adjusted to the observed data on the bottomonium and charmonium. The procedure
for the determination of constituent masses of light and strange quarks with
the given potential is described and tested with the data on the heavy-light
mesons. Then we introduce the spin-dependent forces treated as perturbations.

\subsection{The static potential}

In QCD the static potential is defined in a manifestly gauge invariant way by
means of the vacuum expectation value of a Wilson loop \cite{Su},
\begin{eqnarray}
\label{def_WL}
V(r) &=& - \lim_{T\rightarrow\infty} 
\frac{1}{iT}\, \ln \langle{\cal W}_\Gamma\rangle \;, \nonumber\\
{\cal W}_\Gamma &=& \widetilde{\rm tr}\, 
{\cal P} \exp\left(ig \oint_\Gamma dx_\mu A^\mu\right) \;.
\end{eqnarray}
Here, $\Gamma$ is taken as a rectangular loop with time extension $T$ and
spatial extension $r$. The gauge fields $A_\mu$ are path-ordered along the
loop, while the color trace is normalized according to $\widetilde{\rm
tr}(..)={\rm tr}(..)/{\rm tr}1\!\!1\,$. This definition corresponds to the
calculation of effective action for the case of two external sources fixed at a
distance $r$ during an infinitely long time period $T$, so that the
time-ordering coincides with the path-ordering. Moreover, the contribution into
the effective action by the path parts, where the charges have been separated
to the finite distance during a finite time, can be neglected in comparison
with the infinitely growing term of $V(r)\cdot T$. 

Generally, in the momentum space one rewrites the above definition of QCD
potential of static quarks in terms of relevant quantity $\alpha_{\Rsub V}$
representing a so-called V scheme of QCD coupling constant as follows:
\begin{equation}
 V({\boldsymbol q}^2)  =  -C_F\frac{4\pi\alpha_{\Rsub V}({\boldsymbol
 q}^2)}{{\boldsymbol q}^2}.
\end{equation}

After the introduction of ${\mathfrak a} =\frac{\alpha}{4 \pi}$, the $\beta$
function is actually defined by 
\begin{equation}
\frac{d {\mathfrak a}(\mu^2)}{d\ln\mu^2} = \beta({\mathfrak a})
  = - \sum_{n=0}^\infty \beta_n \cdot {\mathfrak a}^{n+2}(\mu^2),
\end{equation}
so that in the perturbative limit at high ${\boldsymbol q}^2$ the coefficients
were calculated up to three-loop order in the V scheme by the two-loop matching
of static potential with $\alpha_s^{\overline{\Rsub MS}}$ \cite{Peter,Schroed}.
Then $\beta_{0,1}^{\Rsub V}=\beta_{0,1}^{\overline{\Rsub MS}}$ and
$\beta_2^{\Rsub V} = \beta_2^{\overline{\Rsub MS}}-a_1\beta_1^{\overline{\Rsub
MS}} + (a_2-a_1^2)\beta_0^{\overline{\Rsub MS}}$, where the coefficients $a_i$
corresponds to the short-distance expansion
\begin{eqnarray}
\alpha_{\Rsub V}({\boldsymbol q}^2)  &=&  \alpha_{\overline{\Rsub MS}}(\mu^2)
\sum_{n=0}^2 \tilde{a}_n(\mu^2/{{\boldsymbol q}^2})
\left(\frac{\alpha_{\overline{\Rsub MS}}(\mu^2)}{4\pi}\right)^n
= \nonumber\\ &=& 
\alpha_{\overline{\Rsub MS}}({\boldsymbol q}^2)
\sum_{n=0}^2 a_n\left(\frac{\alpha_{\overline{\Rsub MS}}({\boldsymbol q}^2)}
{4\pi}\right)^n. 
\label{vms}
\end{eqnarray}
Note that expansion (\ref{vms}) 
cannot be straightforwardly extended to higher orders of perturbative QCD
because of infrared problems, that result in nonanalytic terms in the
three-loop perturbative potential as was first discussed by Appelquist, Dine
and Muzinich \cite{Su}.

Buchm\"uller and Tye proposed the procedure for the reconstruction of
$\beta$ function in the whole region of charge variation by the known limits of
asymptotic freedom to a given order in $\alpha_s$ and confinement regime.
Generalizing their method, the $\beta_{\rm PT}$ function found in the framework
of asymptotic perturbative theory (PT) to three loops, is transformed to the
$\beta$ function of effective charge as follows:
\begin{eqnarray}
\displaystyle
\frac{1}{\beta_{\rm PT}({\mathfrak a})} &=& -\frac{1}{\beta_0 {\mathfrak a}^2}
+
\frac{\beta_1+\left(\beta_2^{\Rsub V} - \frac{\beta_1^2}{\beta_0}\right)
{\mathfrak a}}{\beta_0^2 {\mathfrak a}} \Longrightarrow \nonumber \\
\frac{1}{\beta({\mathfrak a})} &=& -\frac{1}{\beta_0 {\mathfrak a}^2 \left(1-
\exp\left[-\frac{1}{\beta_0 {\mathfrak a}}\right]\right)} \nonumber \\ &&+
\frac{\beta_1+\left(\beta_2^{\Rsub V} - \frac{\beta_1^2}{\beta_0}\right)
{\mathfrak a}}{\beta_0^2 {\mathfrak a}}
\exp\left[-\frac{l^2 {\mathfrak a}^2}{2}\right],
\label{KKO}
\end{eqnarray}
where the exponential factor in the second term contributes to the
next-to-next-to-leading order at ${\mathfrak a}\to 0$. This function has the
essential peculiarity at ${\mathfrak a}\to 0$, so that the expansion is the
asymptotic series in ${\mathfrak a}$. At ${\mathfrak a}\to \infty$ the $\beta$
function tends to the confinement limit 
\begin{equation}
\frac{d \alpha_{\Rsub V}({{\boldsymbol q}^2})}{d \ln {\boldsymbol q}^2} \to -
\alpha_{\Rsub V}({{\boldsymbol q}^2}),
\label{lim-c}
\end{equation}
which gives the confinement asymptotics for the static potential at long
distances $r$ as usually represented by the linear potential (see discussion in
ref.\cite{simon})
\begin{equation}
V^{\rm conf}(r) = k\cdot r. \label{conf}
\end{equation}
The construction of (\ref{KKO}) is based on the idea to remove the pole from
the coupling constant at a finite energy, but in contrast to the ``analytic''
approach developed in \cite{analit} and modified in \cite{Nesterenko},  the
``smoothing'' of peculiarity occurs in the logarithmic derivative of charge
related with the $\beta$ function, but in the expression for the charge
itself\footnote{In \cite{Nesterenko} the analytic approach is extended to the
$\beta$ function, too.}.
Indeed, in the one-loop perturbation theory we have got
$$
\frac{d\ln {\mathfrak a}}{d\ln \mu^2} = - \beta_0 {\mathfrak a} = -
\frac{1}{\ln
\frac{\mu^2}{\Lambda^2}},
$$
because of 
$$
{\mathfrak a} = \frac{1}{\beta_0 \ln \frac{\mu^2}{\Lambda^2}},
$$
and the pole can be cancelled in the logarithmic derivative itself, so
that
$$
\frac{d\ln {\mathfrak a}}{d\ln \mu^2} \Rightarrow - \beta_0 {\mathfrak a}
\left(1-\frac{\Lambda^2}{\mu^2}\right) \approx  - \beta_0 {\mathfrak a} 
\left(1- \exp\left[-\frac{1}{\beta_0 {\mathfrak a}}\right]\right).
$$
As we see in the perturbative limit ${\mathfrak a}\to 0$ the deviation in the 
$\beta$ function is exponentially small, and the usual solution for the running
coupling constant is valid.

Equation (\ref{KKO}) can be integrated out, so that implicit representation of
effective charge can be inverted by the iteration procedure, so that well
approximated solution has the form
\begin{equation}
{\mathfrak a}(\mu^2) = \frac{1}{\beta_0 \ln\left(1+\eta(\mu^2)
\frac{\mu^2}{\Lambda^2}\right)},
\label{eff}
\end{equation}
where $\eta(\mu^2)$ is expressed through the coefficients of perturbative
$\beta$ function and parameter $l$ in (\ref{KKO}), which is related to the
slope of Regge trajectories $\alpha^\prime_P$ and the integration constant, the
scale $\Lambda$ \cite{KKO}. Thus, the dimensionless parameter $l$ determines
theoretically unknown ratio of perturbative scale in the QCD coupling constant
$\Lambda$ to the confinement scale involved by the Regge slope.

The slope of Regge trajectories, determining the
linear part of potential, is supposed equal to
$ 
\alpha^\prime_P = 1.04\;{\rm GeV}^{-2},
$ so that in (\ref{conf}) we put the parameter
$
k = \frac{1}{2 \pi \alpha^\prime_P}.
$
We use also the measured value of QCD coupling constant \cite{PDG} and pose
$$
\alpha_s^{\overline{\Rsub MS}}(m_Z^2) = 0.123,
$$
as the basic input of the potential. The transformation into the configuration
space was done numerically in \cite{KKO}, so that the potential is presented
in the form of file in the notebook format of MATHEMATICA system. 

The analysis of quark masses and mass spectra of heavy quarkonia results in the
following values ascribed to the potential approach \cite{KKO}:
\begin{equation}
m_c^{\Rsub V} = 1.468\;{\rm GeV,}\;\;\;
m_b^{\Rsub V} = 4.873\;{\rm GeV.}
\label{mcmb}
\end{equation}
Thus, the spectroscopic characteristics of systems composed of nonrelativistic
heavy quarks are determined in the approach with the static potential described
above.

\subsection{Constituent masses of light and strange quarks}

A fast moving light or strange quark interacting with a static source of gluon
field nonperturbatively emits virtual quarks and gluons forming a quark-gluon
sea or a string. Such the cloud of confined virtual fields with the valence
quark is not a local object, which has some internal excitations, of course.
However, these excitations correspond to instable hybrid states related with
additional $q\bar q$ pair or glueball degrees of freedom in the hadron. We do
not consider such exotics in the present paper. Then, the fast moving light
quark surrounded by the virtual soft and ultrasoft fields with no internal
excitations can be described as a whole object with an effective mass, which we
call a constituent mass $\mu_q$. Since we postulate the nonrelativistic
Schr\"odinger equation, we fix the dispersion law of constituent quark: $E_q =
\frac{{\boldsymbol p}^2}{2 \mu_q}$. Further we present a consistent
determination of $\mu_q$, when the static potential entering the Schr\"odinger
equation is fixed.

\begin{figure}[th]
\setlength{\unitlength}{1mm}
\begin{center}
\begin{picture}(80,65)
\put(3,3){\epsfxsize=80\unitlength \epsfbox{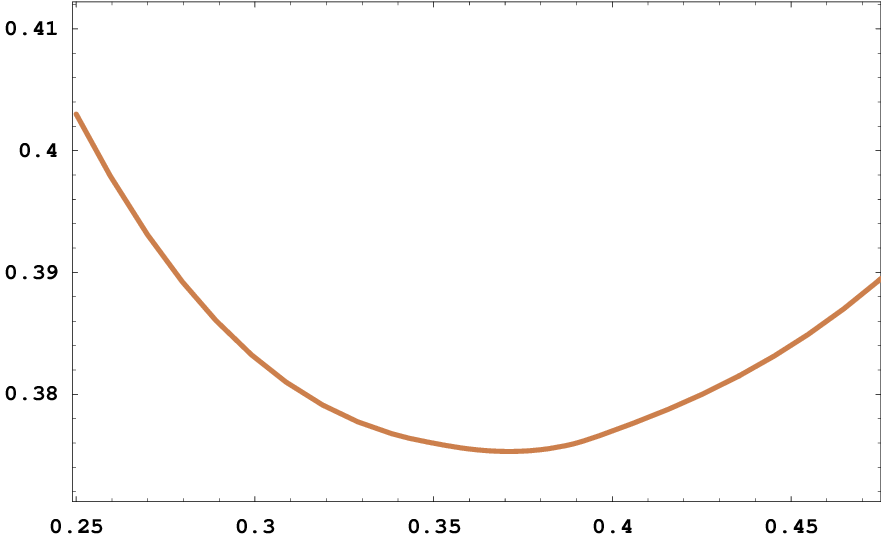}}
\put(70,0){$\mu_q$, GeV}
\put(0,56){$\bar\Lambda(\mu_q)$, GeV}
\end{picture}
\end{center}
\caption{The binding energy of light quark in the potential field of static
source obtained from the solution of (\ref{SE}) versus the constituent mass
$\mu_q$. The normalization $\mu_0=0.37$ GeV corresponds to the minimum of
binding energy.}
\label{flambda}
\end{figure}

The constructive procedure is the following. Since the constituent mass is a
part of energy in the string confining the quarks, and this confining energy is
represented in the static potential by the term linearly growing with the quark
separation, we argue that this part of energy should be subtracted from the
potential. So, we will solve the Schr\"odinger equation for the light quark
with the constituent mass $\mu_q$ and a static source with the potential
$V(r)-\mu_0$, where $\mu_0$ is the energy subtraction. We expect that the
parameters $\mu_q$ and $\mu_0$ are very close to each other, of course. A
difference between them should be suppressed, since it is determined by such
systematic reasons as the nonlocality of constituent quark as well as its
dispersion relation. At fixed $\mu_0$ we can investigate the dependence of
binding energy of constituent quark on the parameter $\mu_q$. So, we
numerically solve the Schr\"odinger equation
\begin{equation}
\left[\frac{{\boldsymbol p}^2}{2 \mu_q} + V(r)\right] \Psi(r) =
[\bar\Lambda(\mu_q)+\mu_0-\mu_q] \Psi(r),
\label{SE}
\end{equation}
for the ground state. Thus, the binding energy is given by
$\bar\Lambda(\mu_q)$, and it is shown in Fig. \ref{flambda}.

We see that the binding energy $\bar\Lambda$ has an optimal value corresponding
to its minimum. We ascribe {\it the position of minimum} $\mu_q=\mu_q^\star
=0.37$ GeV as {\it the valid constituent mass} of light quark attributed to the
Schr\"odinger equation with the given static potential. The difference between
the normalization point $\mu_0$ and the constituent mass $\mu_q^\star$ can be
extracted from the experimental data on the spin-averaged ground-state masses
of heavy-light mesons by comparison with the theoretical expectation
\begin{equation}
M_Q(1S) = m_Q + \left.\bar\Lambda(\mu_q^\star)\right|_{\mu_0=\mu_q^\star} +
\frac{\langle{\boldsymbol
p}^2\rangle}{2 \mu_Q}+\delta\mu,
\label{m1s}
\end{equation}
where we put $\delta\mu =\mu_q^\star-\mu_0$ and use the subleading
approximation including the kinetic energy of heavy quark with a finite mass
$m_Q$, whereas the average $\langle{\boldsymbol p}^2\rangle$ can be estimated
numerically with the wave functions extracted from the Schr\"odinger equation.
Remember, that $\langle{\boldsymbol p}^2\rangle = 2\mu_q^\star\, T$, where $T$
is the kinetic energy, which is phenomenologically independent of the quark
flavor, since the bound states are posed in the intermediate region with the
change of coulomb regime to the linear confinement, so that the potential is
close to the logarithmic form, in which the kinetic energy is
flavor-independent \cite{RQ}.

The experimental data determine the spin-average masses in accordance with
$$
M_Q(1S) = \frac{3 M_V+M_P}{4},
$$
where $M_{V,P}$ are the masses of vector and pseudoscalar states. Then, we find
that (\ref{m1s}) is consistent with the experimental data if
\begin{equation}
\delta\mu = 35\;{\rm MeV,}
\end{equation}
and we see that $\delta\mu\ll\mu_q^\star$ as expected. 

We remark that the same result on $\delta\mu$ can be reproduced by the
numerical solution of relevant Schr\"odinger equation for the two-body problem
with the corresponding kinetic terms for the heavy and light quarks, so that
the difference between the values of $\delta\mu$ in these two approaches occurs
less than 5 MeV, that points to the reliability of this result.

The situation with the determination of constituent mass of strange quark is
very similar, but slightly complicated. Indeed, the strange quark has a
valuable current mass depending on the normalization point. For definiteness we
fix this mass at a low virtuality relevant to the dynamics of bound states, so
that we put $m_s =0.24$ GeV, that is consistent with the extraction of strange
quark mass from the QCD sum rules at the scale $1$ GeV \cite{Narison}. Then we
solve eq.(\ref{SE}) with the substitutions $\mu_q \to \mu_s = m_s+\mu$ and
$\mu_0\to \mu_0^s$, where $\mu$ should be close to $\mu_0^s$, and add the
following
perturbation:
%\begin{equation}
%\delta V_s = k\cdot \left(\langle r\rangle - \frac{\mu_s}{k}\right)\,
%\theta\left(\frac{\mu_s}{k}-\langle r\rangle \right),
%\end{equation}
%or
\begin{equation}
\delta V_0 = k\cdot \left(\langle r\rangle - \langle r_0\rangle\right)\,
\theta\left(\frac{\mu_s}{k}-\langle r\rangle \right),
\label{dV}
\end{equation}
where $k$ is the coefficient in the linear term of potential, the average size
$\langle r\rangle $ is calculated under the wave-functions dependent of the
scale $\mu_s$, so that $\langle r\rangle = \left[\int \Psi^\dagger(r)\Psi(r)
r^2d^3 r\right]^{1/2}$, and this perturbation has quite a clear origin. Indeed,
the length of string having the weight $\mu_s$ equals $\mu_s/k$, and we ascribe
this weight to the constituent mass. However, in the case of quark possessing a
valuable current mass, the hadron can have the size $\langle r\rangle $, which
is less than the attributed length of string $\mu_s/k$ determining the
constituent mass. Then, we have to subtract a part of fake energy that is given
by the excess of length in order to get a consistency of such the
presentation. The value of $\langle r_0\rangle$ is determined at a
normalization point $\mu_0^s$ given by the minimum of binding energy for such
the constituent strange quark.

\begin{figure}[th]
\setlength{\unitlength}{1mm}
\begin{center}
\begin{picture}(80,60)
\put(3,3){\epsfxsize=8cm \epsfbox{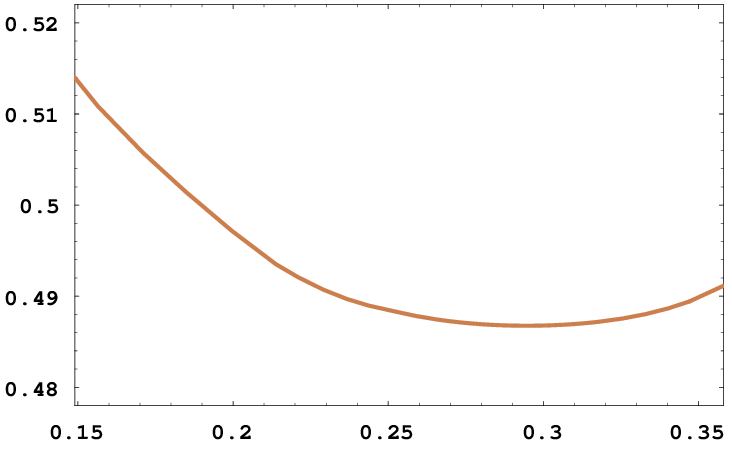}}
\put(65,0){$\mu_s-m_s$, GeV}
\put(0,56){$\bar\Lambda(\mu_s)$, GeV}
\end{picture}
\end{center}
\caption{The binding energy of strange quark in the potential field of static
source obtained from the solution of Schr\"odinger equation versus the
constituent mass $\mu_s$. The normalization $\mu_0^s=0.29$ GeV corresponds to
the minimum of binding energy.}
\label{flambdas}
\end{figure}
The dependence of binding energy in the field of static potential for the
strange quark versus the constituent mass is presented in Fig. \ref{flambdas}.

Thus, in the potential approach we put the position of minimum for the binding
energy as the constituent mass of strange quark, so that
\begin{equation}
\mu_s^\star = 0.53\;{\rm GeV.}
\end{equation}
Following the same procedure as for the light quark we can use the experimental
data on the heavy-strange mesons in order to extract the value of normalization
point $\mu_0^s$ entering as the difference $\delta\mu_s = \mu_s^\star-\mu_0^s$
into the equation of ground state mass, so that using eq.(\ref{m1s}) with
$\langle \boldsymbol p^2\rangle = 2 T \mu_s^\star m_Q/(\mu_s^\star+m_Q)$, we
find
\begin{equation}
\delta\mu_s = 15\;{\rm MeV,}
\end{equation}
while the solution of Schr\"odinger equation taking into account the finite
masses of heavy quarks in $D_s$ and $B_s$ mesons results in 
\begin{equation}
\delta\mu_s = 10\;{\rm MeV,}
\end{equation}
which we will use in our numerical estimates.

The above values of $\delta\mu$ and $\delta\mu_s$ give the estimates of
systematic accuracy of potential model under consideration, so that the
uncertainty of our calculations for the masses of bound states is about $\delta
M = 40$ MeV. 

\begin{figure}[th]
\setlength{\unitlength}{1mm}
\begin{center}
\begin{picture}(80,60)
\put(3,3){\epsfxsize=8cm \epsfbox{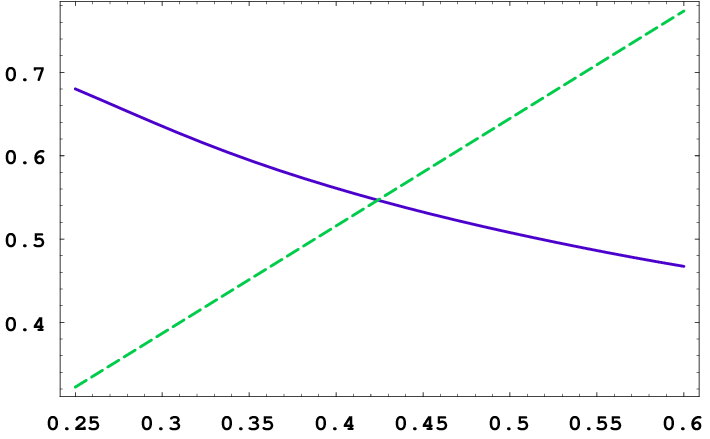}}
\put(65,0){$\mu_s$, GeV}
\put(0,56){$\langle r(\mu_s)\rangle$, $\displaystyle\frac{\mu_s}{k}$, fm}
\end{picture}
\end{center}
\caption{The average size of heavy-strange meson versus the constituent
mass $\mu_s$ (the solid curve). The corresponding length of string with the
weight $\mu_s$ is also show by the dashed line.}
\label{r}
\end{figure}

The difference between the conditions of constituent mass formation in the
heavy-light and heavy-strange mesons is demonstrated in Fig. \ref{r}, where we
show the dependence of average size of meson on the constituent mass in
comparison with the length of string with the weight of constituent mass. We
see that if the constituent mass is less than $0.42$ GeV, then we deal with the
conditions in the heavy-light meson, while at $\mu_s$ greater than $0.42$ GeV,
the situation with the strange quark takes place, since at low virtualities in
the bound state the running mass of strange quark expanded in perturbative
series over $\alpha_s$ has a significant contribution of renormalon
\cite{PhysRepBeneke}, reflecting the infrared singularity in $\alpha_s$. Then,
the running mass of strange quark is greater than $0.42$ GeV, and the
correction (\ref{dV}) is justified. The renormalon contribution can be
subtracted from the running mass, and this can be done by a redefinition of
$\delta\mu_s$ in order to include the renormalon contribution in the value
$\mu_0^s$, as we have performed above, so that $\mu_0^s$ is rather small. Thus,
the running mass of strange quark $m_s=0.24$ represents the so-called
subtracted running mass, that is correlated with the small value of $\mu_0^s$,
since we have rearranged the soft contribution between these two quantities.

Next, the above constituent masses of light and strange quarks are fixed in the
procedure for the ground states of mesons with the static heavy quark, and we
do not perform the same procedure for the excitations and use the fixed values
in the predictions for the doubly heavy baryons, not only for the ground
states, but also for excitations, too.

So, we numerically solve the following Schr\"odinger equations
\begin{eqnarray}
\left[\frac{{\boldsymbol p}^2}{2 \mu_s^\star}+\frac{{\boldsymbol p_{QQ'}}^2}{2
m} +
V(r)\right]&& \Psi_s(r) = \epsilon_s \Psi_s(r), 
\label{SE1}\\[3mm]
\left[\frac{{\boldsymbol p_Q}^2}{2 m_Q}+\frac{{\boldsymbol p_{Q'}}^2}{2 m_{Q'}}
+ \frac{1}{2}V(r)\right] && \Psi_{QQ'}(r) = \epsilon_{QQ'} \Psi_{QQ'}(r), 
\nonumber \label{SE2}
\end{eqnarray}
where the mass of diquark is determined by 
$$
m = m_Q +m_{Q'} + \epsilon_{QQ'},
$$
and the baryon mass is equal to
$$
M = m + m_s + \delta\mu_s + \epsilon_{s}.
$$

Thus, we completely determine the method for the calculation of
spin-independent levels in the systems with the heavy, light and strange
quarks.

\subsection{Spin-dependent corrections}

Following \cite{16,PM1}, we introduce a specified form of spin-dependent
corrections causing the splitting of $nL$-levels, so that in the system of
heavy diquark containing the identical quarks we have
\begin{widetext}
\begin{eqnarray}
V_{SD}^{(d)}({\boldsymbol r}) &=& \frac{1}{2}\,\frac{\boldsymbol L_d\cdot
\boldsymbol S_d}{2m_Q^2}\,
\left( -\frac{dV(r)}{rdr}+
\frac{8}{3}\,\frac{\alpha_s}{r^3}\right) 
\label{SDQQ}\\ %\nonumber \\
&& +\frac{2}{3}\,\frac{\alpha_s}{m_Q^2}\frac{\boldsymbol L_d\cdot \boldsymbol
S_d}{r^3}+\frac{4}{3}\,\frac{\alpha_s}{3m_Q^2}\,{\boldsymbol S}_{Q_1}\cdot
{\boldsymbol S}_{Q_2}\,4\pi\delta({\boldsymbol r}) %\\ && 
-\frac{1}{3}\,\frac{\alpha_s}{m_Q^2}\frac{1}{4\boldsymbol L_d^2 -3} [
6({\boldsymbol L_d\cdot \boldsymbol S_d})^2+3({\boldsymbol L_d\cdot \boldsymbol
S_d})-2\boldsymbol L_d^2\boldsymbol S_d^2]\frac{1}{r^3},
\nonumber     
\end{eqnarray}
\end{widetext}
where the first string corresponds to the relativistic correction to the
effective {\it scalar} exchange, while the second string represents the terms
due to the single-gluon {\it vector} exchange with an effective coupling
constant $\alpha_s$ depending on flavors of quarks composing the system. In
(\ref{SDQQ}) we take into account the color factor corresponding to the
antitriplet state of diquark, i.e. we add $1/2$ in front of usual expression
for the quark-antiquark colorless state, and substitute the static potential
$V(r)$ for the color singlet sources. The last term in (\ref{SDQQ}) represents
the tensor forces expressed in terms of orbital and summed spin of quarks as
was shown in \cite{PRDBc}.

Taking into account the interaction with the strange constituent
quark can be done in an analogous way. So, we have to explore the following
evident kinematics for the motion of two heavy quarks posed in the same point
with the distance $\boldsymbol r$ to the strange quark:
$$
\boldsymbol S_Q \cdot [\boldsymbol r \times \boldsymbol p_Q] = - \frac{1}{2}\,
\boldsymbol S_Q
\cdot \boldsymbol L , \;\;\;\;\;
\boldsymbol S_Q \cdot [\boldsymbol r \times \boldsymbol p_s] = \boldsymbol S_Q
\cdot \boldsymbol L ,
$$
$$
\boldsymbol S_s \cdot [\boldsymbol r \times \boldsymbol p_Q ]=  - \frac{1}{2}\,
\boldsymbol S_s
\cdot \boldsymbol L , \;\;\;\;\;
\boldsymbol S_s \cdot [\boldsymbol r \times \boldsymbol p_s ]=  \boldsymbol S_s
\cdot \boldsymbol L .
$$
In this kinematics, the first term appears in the interaction with the
effective scalar exchange, the second stands in the exchange by the effective
gluon field, and similar expressions appear in the terms with the spin of
strange quark.

Then we adopt the current-current form of interaction between the strange and
heavy quarks with the appropriate antitriplet-state factor of $1/2$ in front of
static potential and sum up the terms related with the heavy-strange
subsystems, so that we can explore a usual technique for the derivation of
spin-dependent perturbations similar to the Breit potential in QED, and for the
interaction of $S$-wave diquark with
the strange quark we get
\begin{widetext}
\begin{eqnarray}
V_{SD}^{(s)}({\boldsymbol r}) &=& \frac{1}{4}\left(\frac{\boldsymbol L\cdot
\boldsymbol S_d}{2m_Q^2} + \frac{4\boldsymbol L\cdot \boldsymbol S_s}
{2{\mu_s^\star}^2}\right) \left( -\frac{dV(r)}{rdr}+
\frac{8}{3}\,\frac{\alpha_s}{r^3}\right) %\nonumber \\ &&
 +\frac{2}{3}\,\frac{\alpha_s}{m_Q \mu_s^\star}\frac{\boldsymbol L\cdot
%%\boldsymbol S_d + 2\boldsymbol L\cdot \boldsymbol S_s}{r^3}+ 
\boldsymbol S}{r^3}+ 
\frac{4}{3}\,\frac{\alpha_s}{3m_Q \mu_s^\star}{{\boldsymbol S_d}
%+{\boldsymbol L_d})
\cdot {\boldsymbol S_s}}\, 4\pi\delta({\boldsymbol r})
\label{SDQs}\\
&& -\frac{1}{3}\,\frac{\alpha_s}{m_Q \mu_s^\star}\frac{1}{4{\boldsymbol L}^2
-3} \left[
6({\boldsymbol L\cdot \boldsymbol S})^2+3({\boldsymbol L\cdot \boldsymbol S})
-2{\boldsymbol L}^2{\boldsymbol S}^2 %\nonumber \\ &&
-6({\boldsymbol L\cdot \boldsymbol S_d})^2-3({\boldsymbol L\cdot \boldsymbol
S_d}) +2{\boldsymbol L}^2\boldsymbol S_d^2\right]
\frac{1}{r^3}, \nonumber    
\end{eqnarray}
\end{widetext}
where ${\boldsymbol S} = {\boldsymbol S_d}+{\boldsymbol S_s}$. In (\ref{SDQs})
we see that this form of spin-dependent forces coincides with the expression
that could be derived under assumption of local doubly heavy diquark with the
spin $\boldsymbol S_d$ interacting with the strange quark, so that the
perturbation in the quark-antiquark system with the diquark mass $m_{QQ} = 2
m_Q$ is exactly reproduced.

The value of effective parameter $\alpha_s$ can be determined by
\begin{equation} 
\alpha_s = \frac{4\pi}{\beta_0\cdot\ln (2\langle T\rangle m_{\rm
red}/\Lambda_{QCD}^2)}, 
\end{equation} 
where $\beta_0 = 11 -2n_f/3$ and $n_f = 3$, $m_{\rm red}$ is
the reduced mass of quarks composing the two-particle system, $T$ is the
kinetic energy in the quark system, so that numerically we get
$\Lambda_{QCD}\approx 113$ MeV from the comparison of theoretical expression
\begin{equation}
\Delta M(ns) = \frac{8}{9}\,\frac{\alpha_s}{m_1m_2}|R_{nS}(0)|^2,
\end{equation}
with the experimental data on the system of $c\bar c$ 
\begin{equation} 
\Delta M(1S,c\bar c) = 117\pm 2\; {\rm MeV,}
\end{equation} 
where $R_{nS}(r)$ is the radial wave function of quarkonium, and it is
calculated in the potential model under study.

In the above estimates we explore the fact that the average kinetic energy of
quarks in the bound state practically does not depend on the flavors of quarks,
and it is given by the following values
\begin{equation}
\langle T_{d}\rangle \approx 0.19\; {\rm GeV,}
\end{equation}
and
\begin{equation}
\langle T_{s}\rangle \approx 0.38\; {\rm GeV,}
\end{equation}
for the antitriplet and singlet color states, correspondingly. 

For the identical quarks inside the diquark, the scheme of $LS$-coupling well
known for the corrections in the heavy quarkonium, is applicable. Otherwise,
for the interaction with the strange quark we use the scheme of $jj$-coupling.
Then, ${\boldsymbol L\cdot \boldsymbol S_s}$ is diagonal at the given
${\boldsymbol J_s}$,  $({\boldsymbol J_s} = {\boldsymbol L} + {\boldsymbol
S_s}$, ${\boldsymbol J} = {\boldsymbol J_s} + \boldsymbol {\bar J})$, where
$\boldsymbol J$ denotes the total spin of baryon, and $\boldsymbol {\bar J}$ is
the total spin of diquark, $\boldsymbol {\bar J}={\boldsymbol S_d}+{\boldsymbol
L_d}$.

In order to estimate various terms and mixings of states, we use
the transformations of bases 
\begin{eqnarray}
|J;J_s\rangle  &=& \sum_{S} |J;S\rangle \,(-1)^{(\bar
J+S_s+L+J)}\times\nonumber\\
&&\sqrt
{(2S+1)(2J_s+1)}
\left\{\begin{array}{ccc} \bar J & S_s & S \\
                        L & J & J_s \end{array}\right\},
\end{eqnarray}
and 
\begin{eqnarray}
|J;J_s\rangle  &= & \sum_{J_d} |J;J_d\rangle \,(-1)^{(\bar
J+S_s+L+J)}\times\nonumber\\
&&
\sqrt {(2J_d+1)(2J_s+1)}
\left\{\begin{array}{ccc} \bar J & L & J_d \\
                        S_s & J & J_s \end{array}\right\},
\end{eqnarray}
where ${\boldsymbol S} = {\boldsymbol S_s} + \boldsymbol {\bar J}$, and
${\boldsymbol J_d} = {\boldsymbol L} + \boldsymbol {\bar J}$.

For the hyper-fine spin-spin splitting in the system of quark-diquark, we use
the presentation with the local diquark for both the interaction of $S$-wave
diquark with the strange quark and that of $P$-wave diquark with the $s$-wave
strange quark. Then we introduce the perturbation analogous to the spin-spin
term in (\ref{SDQs}) with the substitution of $\boldsymbol S_d\to \boldsymbol
{\bar J}$.

Thus, we have defined the procedure of calculations for the mass spectra of
doubly heavy baryons presented in the next Section.

\section{Numerical results}
In this Section we calculate the mass spectra with account for the
spin-dependent splitting of levels. As we have clarified in the Introduction,
the doubly heavy baryons with the identical heavy quarks allow quite a reliable
interpretation in terms of diquark quantum numbers (the summed spin and the
orbital momentum). Dealing with the excitations of $bc$-diquark, we show the
results on the spin-dependent splitting of the ground 1S-state, since the
emission of soft gluon breaks the simple classification of levels for the
higher excitations of such diquark.

The quark-diquark model of bound states is the most reliable for the system
with the more heavy quark. Therefore, the calculations for $\Xi_{bb}$ and
$\Omega_{bb}$ are the most accurate, while the corrections due to the finite
size of diquark can be valuable in $\Xi_{cc}$ and $\Omega_{cc}$.

\begin{table*}[th]
\caption{The spectrum of diquark levels without the spin-dependent
splittings: masses and mean-squared radii.}
\begin{center}
\begin{tabular}{|c|c|c|c|c|c|}
\hline
\multicolumn{6}{|c|}{Diquark $bb$}\\
\hline
diquark level  & mass (GeV) &  $\langle r^2\rangle ^{1/2}$ (fm) &
diquark level  & mass (GeV) &  $\langle r^2\rangle ^{1/2}$ (fm) \\
\hline
1S & 9.72 & 0.33 & 2P & 9.93 & 0.54\\
%\hline
2S & 10.01 & 0.69 & 3P & 10.13 & 0.87\\
%\hline
3S & 10.19 & 0.99 & 4P & 10.30 & 1.14\\
%\hline
4S & 10.35 & 1.26 & 5P & 10.44 & 1.39\\
%\hline
5S & 10.49 & 1.50 & 6P & 10.56 & 1.62\\
%\hline
3D & 10.07 & 0.72 & 4D & 10.24 & 1.02 \\
%\hline
5D & 10.38 & 1.28 & 6D & 10.51 & 1.52 \\
%\hline
4F & 10.18 & 0.87 & 5F & 10.33 & 1.14 \\
%\hline
6F & 10.46 & 1.40 & 5G & 10.27 & 1.01 \\
%\hline
6G & 10.41 & 1.28 & 6M & 10.36 & 1.15\\
%\hline
\hline
\multicolumn{6}{|c|}{Diquark $bc$}\\
\hline
1S & 6.45 & 0.48 & 3P & 6.91 & 1.17\\
%\hline
2S & 6.77 & 0.95 & 4P & 7.11 & 1.52\\
%\hline
3S & 6.99 & 1.34 & 3D & 6.82 & 0.97\\
%\hline
2P & 6.67 & 0.75 & 4D & 7.03 & 1.35\\
%\hline
4F & 6.95 & 1.17 & 5F & 7.14 & 1.53\\
%\hline
5G & 7.07 & 1.35 & 6H & 7.17 & 1.51\\
\hline
\multicolumn{6}{|c|}{Diquark $cc$}\\
\hline
1S & 3.13 & 0.58 & 3P & 3.62 & 1.37\\
%\hline
2S & 3.47 & 1.12 & 4P & 3.85 & 1.78\\
%\hline
3S & 3.72 & 1.57 & 3D & 3.52 & 1.14\\
%\hline
2P & 3.35 & 0.88 & 4D & 3.76 & 1.58\\
\hline
\end{tabular}
\end{center}
\label{diQQ}
\end{table*}

\begin{table*}[th]
\caption{The characteristics of radial wave function for the diquarks:
$R_{d(ns)}(0)$ (GeV$^{3/2}$), $R_{d(np)}^{'} (0)$ (GeV$^{5/2}$).}
%\begin{center}
\begin{tabular}{|c|c|c|c|}
\hline
\multicolumn{4}{|c|}{Diquark $bb$}\\
\hline
nL  & $R_{d(ns)}(0)$  & nL &
$R_{d(np)}'(0)$  \\
\hline
1S & 1.345 & 2P & 0.479  \\
%\hline
2S & 1.028 & 3P & 0.539  \\
\hline
\end{tabular}
\begin{tabular}{|c|c|c|c|}
\hline
\multicolumn{4}{|c|}{Diquark $bb$}\\
\hline
nL  & $R_{d(ns)}(0)$  & nL &
$R_{d(np)}'(0)$  \\
\hline
3S & 0.782 & 4P & 0.585  \\
%\hline
4S & 0.681 & 5P & 0.343  \\
\hline
\end{tabular}
%\end{center}
\begin{tabular}{|c|c|c|c|}
\hline
\multicolumn{4}{|c|}{Diquark $bc$}\\
\hline
nL  & $R_{d(ns)}(0)$  & nL &
$R_{d(np)}'(0)$  \\
\hline
1S & 0.722 & 2P & 0.200  \\
%\hline
2S & 0.597 & 3P & 0.330  \\
\hline
%3S & 0.556 & 4P &    \\
%\hline
\end{tabular}
\begin{tabular}{|c|c|c|c|}
\hline
\multicolumn{4}{|c|}{Diquark $cc$}\\
\hline
nL  & $R_{d(ns)}(0)$  & nL &
$R_{d(np)}'(0)$  \\
\hline
1S & 0.523 & 2P & 0.102  \\
%\hline
2S & 0.424 & 3P & 0.155  \\
\hline
\end{tabular}
\label{wQQ}
\end{table*}

In what follows we consequently refer to the diquark and strange quark quantum
numbers: a principal quantum number $n$ and an orbital momentum, $n_d L_d n_s
l_s$. The results on the characteristics of diquarks: the masses, sizes and
wave functions, calculated with the potential K$^2$O \cite{KKO} are very close
to the values estimated in the Buchm\"uller-Tye potential. The estimates are
presented in Tables \ref{diQQ} and \ref{wQQ}. The wave functions and binding
energies of strange quark slightly depend of the diquark mass.

\subsection{$\Omega_{bb}$ baryons}
Denote the shift of level by $\Delta^{(J)}$ marked by the total spin of
baryon $J$. So, for $1S2p$ we have got
\begin{equation}
\Delta^{(\frac{5}{2})} = -10.5~{\rm MeV}.
\end{equation}
The states with the total spin $J = \frac{3}{2}$ (or $\frac{1}{2}$), can have
different values of $J_s$, and, hence, they have a nonzero mixing, when we
perform the calculations in the perturbation theory built over the states with
the definite total momentum $J_s$ of the strange constituent quark. For $J
=\frac{3}{2}$, the mixing matrix equals
\begin{equation}
\left(\begin{array}{cc} -18.8 & -4.7 \\
                        -4.7 & 39.4 \end{array}\right)~{\rm MeV},
\end{equation}
so that the mixing practically can be neglected, and the level shifts are
determined by the values
\begin{eqnarray}
\Delta^{\prime(\frac{3}{2})} =\lambda_1^{\prime } &=& -19.2~{\rm MeV},\\
\Delta^{(\frac{3}{2})} =\lambda_1 &=& 39.8~{\rm MeV}.\nonumber
\end{eqnarray}
with
\begin{eqnarray}
|1S2p(\frac{3}{2}^{\prime })\rangle  &=& -0.997|J_s=\frac{3}{2}\rangle
-0.080|J_s=\frac{1}{2}\rangle ,\\
|1S2p(\frac{3}{2})\rangle  &=& 0.080|J_s=\frac{3}{2}\rangle
-0.997|J_s=\frac{1}{2}\rangle ,
\nonumber
\end{eqnarray}

For $J=\frac{1}{2}$, the mixing matrix has the form
\begin{equation}
\left(\begin{array}{cc} -23.8 & -15.6 \\
                        -15.6 & 14.2 \end{array}\right)~{\rm MeV},
\end{equation}
with the eigen-vectors given by
\begin{eqnarray}
|1S2p(\frac{1}{2}^{\prime })\rangle  &=& -0.941|J_s=\frac{3}{2}\rangle
-0.338|J_s=\frac{1}{2}\rangle ,\\
|1S2p(\frac{1}{2})\rangle  &=& 0.338|J_s=\frac{3}{2}\rangle
-0.941|J_s=\frac{1}{2}\rangle ,
\nonumber
\end{eqnarray}
and the eigen-values equal
\begin{eqnarray}
\Delta^{\prime(\frac{1}{2})} =\lambda_2^{\prime } &=& -29.5~{\rm MeV},\\
\Delta^{(\frac{1}{2})} =\lambda_2 &=& 19.8~{\rm MeV}.\nonumber
\end{eqnarray}
We straightforwardly check, that the difference between the wave functions
as caused by the different masses of diquark subsystem, is unessential, so that
for the $2S2p$-level the splittings are very close to the values calculated
above.

The splitting of diquark, $\Delta^{(J_d)}$, is numerically small:
$|\Delta^{(J_d)}| < 10$ MeV. Such the corrections are unessential up to the
current accuracy of method ($\delta M\approx 30 - 40$ MeV). They can be
neglected for the excitations, whose sizes are less than the distance to the
strange quark, i.e. for the diquarks with small values of principal quantum
number.

The mass spectra of $\Omega_{bb}$ and $\Xi_{bb}$ baryons are compared in Table
\ref{m-bb}, wherein we restrict ourselves by the presentation of S-, P- and
D-wave levels.

\begin{table*}[th]
\caption{The mass spectrum of $\Omega_{bb}$ baryons in comparison with
$\Xi_{bb}$ one.}
\begin{center}
\begin{tabular}{|l|c|c|l|c|c|}
\hline
$(n_d L_d n_l L_l)$ $J^{P}$ & $M[\Omega_{bb}]$, GeV & $M[\Xi_{bb}]$, GeV
\cite{PM1}
 & $(n_d L_d n_l L_l)$ $J^{P}$ & $M[\Omega_{bb}]$, GeV & $M[\Xi_{bb}]$, GeV
 \cite{PM1}
 \\
\hline
(1S 1s)$1/2^{+}$ & 10.210&10.093 & (3P 1s)$1/2^{-}$&10.617    & 10.493 \\
%\hline
(1S 1s)$3/2^{+}$ & 10.257&10.113  & (3D 1s)$5/2^{\prime +}$ & - & 10.497 \\
%\hline
(2P 1s)$1/2^{-}$&10.416 & 10.310  & (3D 1s)$7/2^{+}$&10.627   & 10.510 \\
%\hline
(2P 1s)$3/2^{-}$ &10.462& 10.343 & (3P 1s)$3/2^{-}$ &10.663  & 10.533 \\
%\hline
(2S 1s)$1/2^{+}$ & 10.493 &10.373& (1S 2p)$1/2^{-}$ &10.651& 10.541  \\
%\hline
(2S 1s)$3/2^{+}$ & 10.540&10.413 & (1S 2p)$3/2^{-}$&10.661 & 10.567 \\
%\hline
(3D 1s)$5/2^{+}$& - & 10.416  & (1S 2p)$1/2^{\prime -}$&10.700 & 10.578 \\
%\hline
(3D 1s)$3/2^{\prime +}$& - & 10.430 & (1S 2p)$5/2^{-}$ &10.670 & 10.580 \\
%\hline
(3D 1s)$1/2^{+}$&10.617 & 10.463 & (1S 2p)$3/2^{\prime -}$ &10.720& 10.581  \\
%\hline
(3D 1s)$3/2^{+}$& - & 10.483 & (3S 1s)$1/2^{+}$   &  10.682&10.563 \\
\hline
\end{tabular}
\end{center}
\label{m-bb}
\end{table*}

The most reliable predictions are the masses of baryons $1S1s\; (J^P=3/2^+,\;
1/2^+)$, $2P1s\; (J^P=3/2^-,\; 1/2^-)$ and $3D1s\; (J^P=7/2^+,\ldots 1/2^+)$.
The $2P1s$-level is quasistable. In the $\Xi_{bb}$ family the transition into
the ground state requires the instantaneous change of both the orbital momentum
and the summed spin of quarks inside the diquark. Therefore,  when the
splitting between $2P1s$ and $1S2p$, $\Delta E\sim \Lambda_{QCD}$, is not
small, their mixing is suppressed as $\delta V/\Delta E \sim \frac{1}{m_Q m_q}
\frac{r_d}{r_s^4} \frac{1}{\Delta E} \ll 1$. Since the admixture of $1S2p$ in
the $2P1s$-state is low, the  $2P1s$-levels are quasistable, i.e. their
hadronic transitions into the ground state with the emission of $\pi$-mesons 
are suppressed as we have derived, though an additional suppression is given by
a small value of phase space. In contract to the $\Xi_{bb}$ family, in the
$\Omega_{bb}$ system the transition of $2P1s$ to the ground level with the
emission of pion is forbidden because of the conservation of flavor in the
strong interactions, while the emission of kaon with the transition in to the
ground state of $\Xi_{bb}$ is forbidden by nonsufficient phase space. An
alternative possibility is the transition under the emission of two pions in
the singlet of the isospin group, that is the open channel for the $2S1s$
levels.

As for the higher excitations, the $3P1s$-states are close to the $1S2p$-levels
with $J^P=3/2^-,\; 1/2^-$, so that the operators changing both the
orbital momentum of diquark and its spin, can lead to the essential mixing with
an amplitude $\delta V_{nn'}/\Delta E_{nn'}\sim 1$, despite of suppression by
the inverse heavy quark mass and small size of diquark. The mixing slightly
shifts the masses of states. The most important effect is a large admixture of
$1S2p$ in $3P1s$. It makes the state to be unstable because of the transition
into the ground $1S1s$-state with the emission of gluon (the E1-transition).
This transition leads to decays with the emission of kaons\footnote{Remember,
that the $\Xi_{QQ'}$-baryons are the iso-dublets, while the $\Omega_{QQ'}$ ones
are the iso-singlets.}. 

The level $1S2p\; J^P=5/2^-$ has the definite quantum numbers of diquark and
light quark motion, because there are no levels with the same values of $J^P$
in its vicinity. However, its width of transition into the ground state of
$\Xi_{bb}$ and kaon is not suppressed and seems to be large, $\Gamma\sim 100$
MeV.

One could expect the transitions of 
$$
\Omega_{bb}\left({\frac{3}{2}}^-\right) \to \Xi_{bb}\left({\frac{3}{2}}^+
\right) K \;\; {\rm in~S-wave,}
$$
$$
\Omega_{bb}\left({\frac{3}{2}}^-\right) \to \Xi_{bb}\left(\frac{1}{2}^+
\right) K \;\; {\rm in~D-wave,}
$$
$$
\Omega_{bb}\left({\frac{1}{2}}^-\right) \to \Xi_{bb}\left({\frac{3}{2}}^+ 
\right) K \;\; {\rm in~ D-wave,}
$$
$$
\Omega_{bb}\left({\frac{1}{2}}^-\right) \to \Xi_{bb}\left({\frac{1}{2}}^+ 
\right) K\;\; {\rm in~ S-wave.}
$$
The D-wave transitions are suppressed by the ratio of low recoil momentum to
the mass of baryon.

The width of $1S1s$ state with $J^P=3/2^+$ is completely determined by the
radiative electromagnetic M1-transition into the basic $J^P=1/2^+$ state.

\subsection{$\Omega_{cc}$ baryons}
The calculation procedure described above leads to the results for the
doubly charmed baryons as presented below.

For $1S2p$, the splitting is equal to
\begin{equation}
\Delta^{(\frac{5}{2})} = 5.1~{\rm MeV}.
\end{equation}
For $J =\frac{3}{2}$, the mixing is determined by the matrix
\begin{equation}
\left(\begin{array}{cc} -18.5 & -14.8 \\
                        -14.8 & 42.9 \end{array}\right)~{\rm MeV,}
\end{equation}
so that the eigen-vectors
\begin{eqnarray}
|1S2p(\frac{3}{2}^{\prime })\rangle  &=& -0.975|J_s=\frac{3}{2}\rangle
-0.223|J_s=\frac{1}{2}\rangle ,\\
|1S2p(\frac{3}{2})\rangle  &=& 0.223|J_s=\frac{3}{2}\rangle
-0.975|J_s=\frac{1}{2}\rangle ,
\nonumber
\end{eqnarray}
have the eigen-values
\begin{eqnarray}
\Delta^{\prime(\frac{3}{2})} =\lambda_1^{\prime } &=& -21.9~{\rm MeV},\\
\Delta^{(\frac{3}{2})} =\lambda_1 &=& 46.3~{\rm MeV}.\nonumber
\end{eqnarray}
For $J=\frac{1}{2}$, the mixing matrix equals
\begin{equation}
\left(\begin{array}{cc} -32.6 & -47.6 \\
                        -47.6 & -31.4 \end{array}\right)~{\rm MeV,}
\end{equation}
where the vectors
\begin{eqnarray}
|1S2p(\frac{1}{2}^{\prime })\rangle  &=& -0.712|J_s=\frac{3}{2}\rangle
-0.703|J_s=\frac{1}{2}\rangle ,\\
|1S2p(\frac{1}{2})\rangle  &=& 0.703|J_s=\frac{3}{2}\rangle
-0.712|J_s=\frac{1}{2}\rangle ,
\nonumber
\end{eqnarray}
\begin{table*}[th]
\caption{The mass spectrum of $\Omega_{cc}$ baryons in comparison with
$\Xi_{cc}$ one.}
\begin{center}
\begin{tabular}{|l|c|c|l|c|c|}
\hline
$(n_d L_d n_l L_l)$ $J^{P}$ & $M[\Omega_{cc}]$, GeV & $M[\Xi_{cc}]$, GeV
\cite{PM1}
 & $(n_d L_d n_l L_l)$ $J^{P}$ & $M[\Omega_{cc}]$, GeV & $M[\Xi_{cc}]$, GeV
 \cite{PM1}
 \\
\hline
(1S 1s)$1/2^{+}$ &3.594& 3.478 & (3P 1s)$1/2^{-}$&4.073  & 3.972 \\
%\hline
(1S 1s)$3/2^{+}$ &3.730& 3.61  & (3D 1s)$3/2^{\prime +}$ & - & 4.007 \\
%\hline
(2P 1s)$1/2^{-}$ &3.812& 3.702  & (1S 2p)$3/2^{\prime -}$ &4.102& 4.034  \\
%\hline
(3D 1s)$5/2^{+}$ & - & 3.781  & (1S 2p)$3/2^{-}$ &4.176 &4.039 \\
%\hline
(2S 1s)$1/2^{+}$ &3.925& 3.812 & (1S 2p)$5/2^{-}$  &4.134& 4.047\\
%\hline
(3D 1s)$3/2^{+}$ & - & 3.83 & (3D 1s)$5/2^{\prime +}$ & - & 4.05 \\
%\hline
(2P 1s)$3/2^{-}$ &3.949& 3.834 & (1S 2p)$1/2^{\prime -}$&4.145 & 4.052 \\
%\hline
(3D 1s)$1/2^{+}$ &3.973& 3.875 & (3S 1s)$1/2^{+}$ &4.172  & 4.072\\
%\hline
(1S 2p)$1/2^{-}$ &4.050& 3.927  & (3D 1s)$7/2^{+}$  &4.204 & 4.089 \\
%\hline
(2S 1s)$3/2^{+}$ &4.064& 3.944 & (3P 1s)$3/2^{-}$ &4.213  & 4.104 \\
\hline
%           &       & (3S 1s)$3/2^{+}$ & 4.204 \\
%%\hline
\end{tabular}
\end{center}
\label{m-cc}
\end{table*}

\noindent
have the eigen-values
\begin{eqnarray}
\Delta^{\prime(\frac{1}{2})} =\lambda_2^{\prime } &=& -79.6~{\rm MeV},\\
\Delta^{(\frac{1}{2})} =\lambda_2 &=& 15.6~{\rm MeV}.\nonumber
\end{eqnarray}

For the $1S$-, $2S$- and $3S$-wave levels of diquark, the shifts of vector
states are equal to
\begin{eqnarray}
\Delta (1S) &=& 6.4~{\rm MeV},\nonumber\\
\Delta (2S) &=& 4.7~{\rm MeV}.\nonumber\\
\Delta (3S) &=& 4.2~{\rm MeV}.\nonumber
\end{eqnarray}

The mass spectra of the $\Omega_{cc}$ and $\Xi_{cc}$ baryons are
presented in Table \ref{m-cc}.

\subsection{$\Omega_{bc}$ baryons}
As we have already mentioned in the Introduction, the heavy diquark composed of
the quarks of different flavors, turns out to be unstable under the
emission of soft gluons. We suppose that the calculations of masses for the
excited $\Omega_{bc}$ baryons are not so justified in the given scheme.
Therefore, we present only the result for the lowest states with $J^P=1/2^+$
$$
M_{\Omega_{bc}^{\prime}} = 6.97\; {\rm GeV,}\quad M_{\Omega_{bc}} = 6.93\;
{\rm GeV,}\quad 
$$
and with $J^P=3/2^+$
$$
M_{\Omega_{bc}^\ast} = 7.00\; {\rm GeV,}
$$
whereas for the vector diquark we have assumed that the spin-dependent
splitting due to the interaction with the strange quark is determined by the
standard contact coupling of magnetic moments for the point-like systems. 

\section{Conclusion}
In this paper we have evaluated the spectroscopic characteristics of baryons
containing two heavy quarks and the single strange quark, in the framework of
potential model. The calculations have been based on the assumption of
string-like structure of doubly heavy baryon, when the diquark of small size
interacts with the strange quark in the limit of quark-diquark factorization in
the wave functions. We have explored the QCD-motivated model of static
potential \cite{KKO}, that takes into account two known asymptotic regimes at
small and large distances. The first limit is the asymptotic freedom up to
three-loop running of coupling constant consistent with the measurements of
$\alpha_s$ at large virtualities. The second regime is the linearly raising
confining potential. The spin-dependent corrections have been taken into
account. The region of factorization applicability as well as the uncertainties
have been discussed.

Below the threshold of decay into the heavy baryon and heavy strange meson, we
have found the system of excited bound states, which are quasistable under the
hadronic transitions into the ground state. In the baryonic systems with two
heavy quarks and the strange quark, the quasistability of diquark excitations
is provided by the absence of transitions with the emission of both a single
kaon and a single pion. These transitions are forbidden because of small
splitting between the levels and the conservation of the iso-spin and
strangeness. Further studies on the electromagnetic and hadronic transitions
between the states of doubly heavy baryons are of interest.

In conclusion we compare the results obtained in the present paper with the
estimates in potential models and in lattice simulations as shown in Table 
\ref{compar}. 

\begin{table}[th]
\caption{The masses of ground states $M$ (in GeV) for the baryons with two
heavy quarks calculated in various approaches (* denotes the results of authors
in this work). The accuracy of predictions under the variation of model
parameters is about 30-50 MeV. The systematic uncertainties are discussed in
the text.}
\label{compar}
\begin{center}
\begin{tabular}{|c|c|c|c|c|c|c|c|c|c|}    \hline
baryon & *  & \cite{faust} & \cite{guo} & \cite{Ron} & \cite{Korner} &
\cite{itoh} & \cite{kaur} & \cite{faust2} & \cite{RMW}\\
\hline
$\Xi_{cc}$ & 3.48 & 3.66 & 3.74 & 3.66 & 3.61 & 3.65 & 3.71 & 3.62 & 3.57 
\\  \hline
$\Xi_{cc}^\ast$& 3.61 & 3.81 & 3.86 &3.74 & 3.68 & 3.73 & 3.79 & 3.73 & 3.63
\\  \hline
$\Omega_{cc}$ & 3.59 & 3.76 & 3.76 & 3.74 & 3.71 & 3.75 & 3.89 & 3.78 & 3.69
\\   \hline
$\Omega_{cc}^\ast$ & 3.73 & 3.89 & 3.90 & 3.82 & 3.76 & 3.83 & 3.91 & 3.87 &
3.75
\\   \hline
$\Xi_{bb}$ & 10.09 & 10.23 & 10.30 & 10.34 & - & - & 10.43 & 10.20 & -
\\  \hline
$\Xi_{bb}^\ast$& 10.11 & 10.28 & 10.34 & 10.37 & - & - & 10.48 & 10.24 & -
\\  \hline
$\Omega_{bb}$  & 10.21 & 10.32 & 10.34 & 10.37 & - & - & 10.59 & 10.36 & -
\\   \hline
$\Omega_{bb}^\ast$& 10.26  & 10.36 & 10.38 & 10.40 & -  & - & 10.62 & 10.39 & -
\\   \hline
$\Xi_{cb}$     & 6.82 & 6.95 & 7.01 & 7.04 & - & - & 7.08 & 6.93 & 6.84
\\  \hline
$\Xi'_{cb}$    & 6.85 & 7.00 & 7.07 & 6.99 & - & - & 7.10 & 6.96 & 6.83
\\  \hline 
$\Xi_{cb}^\ast$& 6.90 & 7.02 & 7.10 & 7.06 & - & - & 7.13 & 6.98 & 6.88
\\  \hline
$\Omega_{cb}$  & 6.93 & 7.05 & 7.05 & 7.09 & - & - & 7.23 & 7.09 & 6.95
\\  \hline
$\Omega'_{cb}$ & 6.97 & 7.09 & 7.11 & 7.06  & - & - & 7.24 & 7.12 & 6.94
\\  \hline
$\Omega_{cb}^\ast$& 7.00 & 7.11 & 7.13 & 7.12 & -  & - & 7.27 & 7.13 & 6.98
\\  \hline
\end{tabular}
\end{center}
\end{table}

The quark-diquark factorization in calculating the masses of ground states for
the baryon systems with two heavy quarks was also considered in ref.
\cite{faust}, where the quasi-potential approach \cite{quasipot} was explored.
There is a numerical difference in the choice of heavy quark masses, that leads
to that in \cite{faust} the mass of doubly charmed diquark, say, about 100 MeV
greater that the mass used in the above calculations. This difference
determines the discrepancy of estimates for the masses of ground states
presented in this paper and in \cite{faust}. We believe that this deviation
between the quark masses is caused by the use of Cornell potential with the
constant value of effective coulomb exchange coupling in contrast to the above
consideration with the running coupling constant, that cancels the uncertainty
in the arbitrary additive shift of energy. Furthermore, in the potential
approach the masses of heavy quarks depend on the mentioned additive shift,
which adjusted in the phenomemological models by comparing, say, the leptonic
constants of heavy quarkonium calculated in the model with the values known
from experiments. In the QCD motivated potential such the ambiguity of
potential because of the additive shift is absent, so that the estimates of
heavy quark masses have less uncertainties. Let us stress that in the Cornell
model the leptonic constants were calculated by taking into account the
one-loop corrections caused by the hard gluons. This correction is quite
essential, in part, for the charmed quarks. The two-loop corrections are also
important for the consideration of leptonic constants in the potential approach
\cite{KKO}. Moreover, in \cite{faust} the constituent mass of light quark is
posed with no correlation with the normalization of potential, while we put the
constituent mass to be a part of nonperturbative energy in the potential. This
can lead to an additional deviation between the estimates of baryon masses
about 50 MeV. Taking into account the above notes on the systematic
differences, we can claim that the estimates of ground states masses for the
baryons with two heavy quarks in \cite{faust} agree with the values obtained in
the presented approach (see Table \ref{compar}).

In ref. \cite{guo}, following \cite{faust} in the framework of quasi-potential
approach, the analysis of spin-dependent relativistic corrections was performed
so that the overestimated, to our opinion, value of heavy diquark mass from
\cite{faust} was used. Unfortunately, there is an evident mistake in the
description of calculations in \cite{guo}, because both the parameter giving
the relative contribution of scalar and vector part in the potential and the
anomalous chromo-magnetic moment of heavy quark are denoted by the same symbol,
that leads to numerical errors, since in \cite{faust} it was shown that these
quantities have different values. This mistake enlarges the uncertainty about
100 MeV into the estimates of \cite{guo}, so that we can consider that the
results of \cite{guo} do not contradict with the presented description (see
Table \ref{compar}).

The estimates based on the hypothesis of pair interactions were presented in
ref. \cite{Ron}, so that in the light of discussion given in the Introduction
the difference about 200-300 MeV, that follows from values in Table 
\ref{compar}, is not amazing. This deviation is, in general, related with the
different character of interquark forces in the doubly heavy baryon, though the
uncertainty in the heavy quark masses is also important.

In ref. \cite{Korner} simple speculations based on the HQET with the heavy
diquark were explored, so that the estimates depend on the supposed mass of
diquark composed of two heavy quarks. In this way, if we neglected the binding
energy in the diquark, that is evidently related with the choice of heavy quark
masses, then we got the estimates of ground state masses shown in Table
\ref{compar}.

Next, in \cite{kaur} the analysis given in \cite{khanna} was modified on the
basis of interpolation formulae for the mass of ground state with account for
the dependence of spin forces on both the wave functions and the effective
coupling constant, which were changed with the quark contents of hadrons. In
this way, the parameter of energy shift enters the fitting function, so that
this parameter essentially changes under the transition from the description of
mesons to baryons: $\delta_M\approx 80$ MeV $\longrightarrow$ $\delta_B\approx
210$ MeV. This shift of energy provides a good agreement of fitting with the
mass values for the mesons and baryons observed experimentally. However, if we
suggest that the doubly heavy baryon is similar with the meson containing the
local heavy source in the picture of strong interactions, then we should use
the energy shift prescribed to the heavy mesons but the heavy baryons, wherein
the presence of system with two light quarks leads to the essential difference
in the calculation of bound state masses, hence, to the energy shift different
from the mesonic one. Such the substitution of parameters would lead to a more
good agreement between the results of \cite{kaur} (see Table \ref{compar}) and
the values obtained in this work.

Recently the analysis presented in \cite{faust} was modified in \cite{faust2}
in order to take into account the running of QCD coupling in the static
potential in the framework of quasi-potential approach. The authors of
\cite{faust2} claim that the difference between the estimates of masses for
doubly heavy baryons in \cite{PM1} and \cite{faust2} accumulates three sources.
The first is the choice of heavy quark masses, as we have mentioned above, that
gives the shift about 50 MeV by the estimates in \cite{faust2}. The second
source is the relativistic dispersion law of free light or strange quarks. The
third is the spin-independent relativistic corrections for the orbitally
excited states. Nevertheless, we emphasize that the estimates obtained in these
two approaches are in a good agreement within the limits of systematic
uncertainties about 70 MeV.

Finally, the lattice simulations based on the lagrangian of NRQCD are presented
in Table \ref{compar}, too \cite{RMW}. One can see that the lattice results
show approximately twice reduction of spin-dependent splitting of the ground
level, which position agrees with the estimates obtained in the potential
models.

Summarizing, we can claim that, first of all, in the framework of potential
approach in the calculations of masses for the doubly heavy baryons the
dominant uncertainty is caused by the choice of heavy quark masses, so that due
to the adjustment on the systems with heavy quarks, the analysis presented in
the QCD-motivated model of potential with the running coupling constant at
short distances and the linear nonperturbative term confining quarks at large
distances, gives the most reliable predictions. Further, the calculations in
the framework of NRQCD sum rules \cite{QCDSRQQq} gave the results for the
ground state masses with no account of spin-dependent forces, so that the sum
rule estimates are in a good agreement with the values obtained in the 
potential models.

\section*{Acknowledgements}
This work is in part supported by the Russian Foundation for Basic Research,
grants 01-02-99315, 01-02-16585, 02-02-16253-a and 00-15-96645, the Federal
program ``State support for the integration of high education and fundamental
science'', grant 247 (V.V.K. and O.N.P.), and the Russian Ministry of
Education, grant 00-3.3-62.

%\newpage

\end{document}